\documentclass[12pt]{article}

\usepackage{lineno}
\modulolinenumbers[5]

\usepackage{listings}
\usepackage[most]{tcolorbox}
\usepackage{inconsolata}

\newtcblisting[auto counter]{mycode}[2][]{
    fonttitle=\bfseries, colframe=gray, listing only, 
    listing options={basicstyle=\ttfamily,language=java,showstringspaces=false}, 
    title=Example #2: code}

\newtcblisting[auto counter]{myout}[2][]{
    fonttitle=\bfseries, colframe=gray, colback=white, listing only, 
    listing options={basicstyle=\ttfamily,language=java}, 
    title=Example #2: output}

\usepackage{authblk}
\usepackage{float}

\begin{document}

\title{\Large\textbf{BOAT: a cross-platform software for data analysis and numerical computing with arbitrary-precision}}
\author{Davide Pagano}
\affil{Dipartimento di Ingegneria Meccanica e Industriale,\\
 Universit$\grave{a}$ degli Studi di Brescia}
 \date{}   
\renewcommand\Affilfont{\itshape\small}
%\author{\footnote{ }
%  \texttt{davide.pagano@unibs.it}}

\maketitle

\begin{abstract}
BOAT is a free cross-platform software for statistical data analysis and numerical computing. Thanks to its multiple-precision floating point engine, it allows arbitrary-precision calculations, whose digits of precision are only limited by the amount of memory of the host machine. At the core of the software is a simple and efficient expression language, whose use is facilitated by the assisted typing, the auto-complete engine and the built-in help for the syntax.
In this paper a quick overview of the software is given. Detailed information, together with its applications to some case studies, is available at the BOAT web page \cite{BOATWEB}.  
\end{abstract}

%\begin{keyword}
%\texttt{elsarticle.cls}\sep \LaTeX\sep Elsevier \sep template
%\MSC[2010] 00-01\sep  99-00
%\end{keyword}

%\end{frontmatter}

\linenumbers

\section{Introduction}
\label{introduction}

At present, several tools for statistical data analysis and numeric computing, including free and commercial software, are available on the market. Some of them have become industry \textit{standards}, like SAS \cite{SASWEB} and SPSS \cite{SPSSWEB}, while others like R \cite{RWEB}, MATLAB \cite{MATLABWEB} and ROOT \cite{ROOTWEB} are widely used among scientists. 

In this paper a new software, called BOAT, is presented. Obviously, it has not been 
thought and developed as a replacement of the previously mentioned software, which have benefited (feature-wise) from decades of development. Instead, the idea was to create a framework where to perform frequently used statistical analyses (t-test, z-test, ANOVA, normality tests, etc...) and numerical computing with ease (thanks to a very simple and efficient syntax) and with arbitrary precision. The latter feature, in particular, makes BOAT more suitable, than many other available tools, in those scenarios where many digits of precision are requested. This is briefly explained in the following.
\\

At present most of the floating-point computations are performed in double precision. The IEEE 754 standard \cite{1985--ieee754} specifies the binary64 format (corresponding to the double data-types in the ISO C language) as having a significand precision of 53 bits, which corresponds to about 16 decimal digits of precision. Although this precision is more than enough in most of the practical cases, several applications in fields like Physics, Mathematics or Engineering require a higher precision.

Such applications include \cite{math3020337} studies on planetary orbit dynamics, zeroes of the Riemann Zeta function, Ising-class integrals or electromagnetic scattering theory to name a few. Another notable field of application is data security and in particular cryptography algorithms, like for example the RSA \cite{Rivest78amethod}. Its security is based on the very challenging problem of the decomposition of an integer into two prime numbers, when very large primes are involved. High level of security requires to handles prime numbers with many hundreds of digits \cite{Meijer}.  

In order to handle these scenarios, in the last decades, some libraries for arbitrary precision arithmetic have been developed, like the GNU Multiple Precision Arithmetic Library \cite{GMP}. The general idea is to split large significands over several machine words of 64 bits or 32 bits, depending on the system architecture. This allows to reach a number of precision digits which is only limited by the available memory of the host machine. Notwithstanding the availability of these libraries, the number of software for statistical data analysis with full support of 
arbitrary precision arithmetic is quite limited \cite{LIstSoft}.
\\

BOAT allows arbitrary-precision calculations, not only for integer and float numbers, but also for complex numbers, vectors and matrices. Since high precision calculations can (sensibly) impact the performance, BOAT allows the user to choose the precision before each operation. In this way, many digits of precision can be reserved just for those calculations, which could really benefit from them. In addition to that, an automatic downcast to \textit{double} is performed for all those cases where arbitrary precision is not required, like for charts. A more extensive explanation about the arbitrary precision in BOAT is given in Section \ref{arbprec}.
\\

The support for arbitrary-precision calculations is only one of the key features in BOAT. Others include:

\begin{itemize}
\item compatibility with OSX, Windows and Linux (in beta);
\item a modern graphical user interface with retina display support on OSX;
\item a simple and effective expression language;
\item an input console with assisted typing, syntax highlight and auto-complete functions;
\item the support for several statistical analyses;
\item a built-in help system.
\end{itemize}

In the following section an overview of the graphical user interface is given.

\begin{figure}[H]
\label{GUI}
  \centering
  \caption{The graphical user interface of BOAT on OSX.}
  \includegraphics[width=\textwidth]%
    {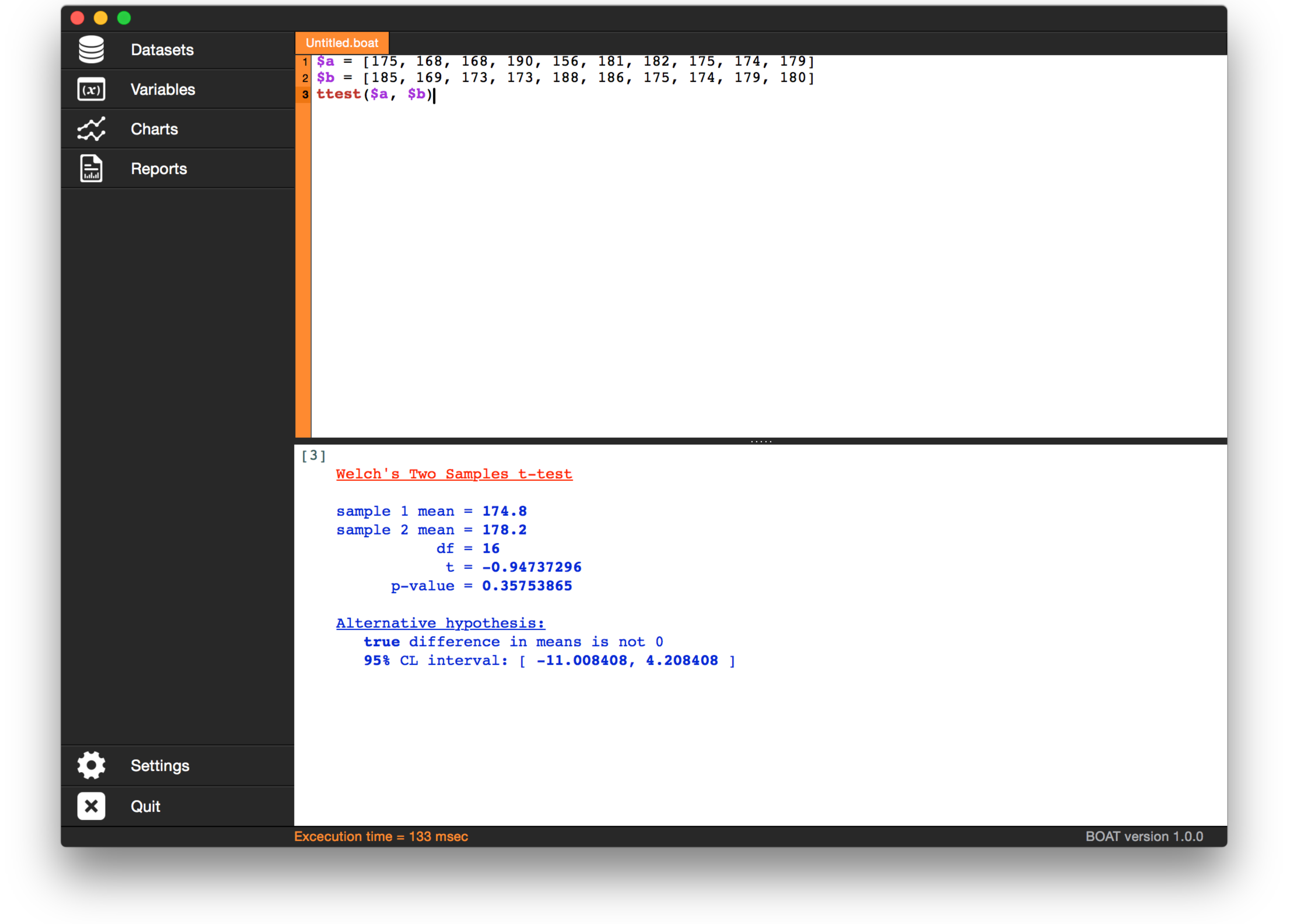}
\end{figure}

\section{The Graphical User Interface}

The main graphical user interface in BOAT is shown in Figure \ref{GUI}. It consists of three main elements: the \textit{info panel} on the left, the \textit{input console} on top-right of the window, and the \textit{output console} which is placed just below the input console. Both input and output consoles are fully resizable. 

\subsection{The info panel}
\label{infopanel}

The info panel, organized into expandable tabs, provides an overview of all the  \textit{objects} which are temporarily or permanently stored in memory or on disk. BOAT defines four types of objects:

\begin{itemize}
\item \textit{dataset}: a collection of data organized as a table of observations (rows) and variables (columns);
\item \textit{variable}: an identifier associated to a value of any supported data type (see Section \ref{datatypes});
\item \textit{chart}: an identifier associated to a graphical representation of data;
\item \textit{report}: any non BOAT-specific object (images, HTML pages, etc...).
\end{itemize}

The info panel also facilitates to manage objects, like delete or recall variables, exporting charts, edit datasets, etc... 

\subsection{The input console}

The input console is the interface between the user and the BOAT expression language, and it has been designed to help writing the code. The list of features includes: syntax highlight, auto-complete, parentheses matching and assisted typing. The input console also supports editing multiple files, which are arranged into different tabs. Projects in different tabs will share all the objects listed in the info panel.

\subsection{The output console}

Most of the operations performed in BOAT result in a text output, which is shown in the output console. The content of the output console, being related to the code execution, is always temporary but it could be converted to a report object (plain text and HTML) to store it on disk permanently.

\section{Syntax Overview}

In this section a quick overview of the syntax is given. More detailed information is available on the online user guide \cite{BOATWEB}.
BOAT is built around a user-friendly expression language by which every construction is an expression that returns an object (number, matrix, chart, etc...). 

Expressions are separated by a newline and basic \textit{statements} in BOAT consist of either an expression evaluation or a variable assignment. In the first case the result of the evaluation is printed on the output console and it is not saved, whereas, in an assignment statement, the result is passed to a variable and it is not displayed. This is clarified in the next example.

In Example 1 three statements are defined. The first line is a numerical calculation with integer numbers: BOAT evaluates the expression and displays the result in the output console (as shown Example 1: output). The second line is a variable assignment, which stores the result (of the same calculation as before) into a variable named \verb|myvar|, and does not produce any output on screen. Finally, the third statement is again an expression evaluation, where the value associated to the previously defined variable is accessed. The corresponding result is again sent to the output console.

\vspace{2mm}
\small
\begin{center}
\begin{minipage}{0.52\textwidth}
\begin{mycode}{1}
2^( 3 + 1 ) / 4
$myvar = 2^( 3 + 1 ) / 4
$MyVar * 3
\end{mycode} 
\end{minipage}%
\hspace{4mm}%
\begin{minipage}{0.38\textwidth}
\begin{myout}{1}
4

12
\end{myout} 
\end{minipage}
\end{center}
\normalsize
\vspace{2mm}

Example 1 also shows two other important features of the expression language in BOAT:

\begin{itemize}
\item variables are identified by the prefix \verb|$| and their data type does not have to be specified, as it is automatically determined;
\item BOAT is \textbf{case insensitive}, so variables like \verb|$myvar| and \verb|$MyVar| are equivalent, and so are \textit{functions} like \verb|sqrt| and \verb|sQRt|.
\end{itemize}

As previously stated, expressions are separated by a newline. However, in some practical cases, it could be preferable to handle a single long expression by splitting it over few lines. This can be done by placing the special char \verb|%| at the begin and end of the expression, as shown in Example 2.

\vspace{2mm}
\small
\begin{mycode}{2}
% plot( [1, 2, 3],[3, -2, 1], 
title = "test plot", 
Xtitle = "x axis", 
Ytitle = "y axis") %
\end{mycode} 
\normalsize
\vspace{2mm}

Expressions can also be composed by one or more \textit{functions} or one \textit{command}. \textit{Functions} in BOAT are operations which act on one or more of the supported data classes (see Section \ref{datatypes}) and always return an object. On the other hand \textit{commands} do not work on data and do not always return an object (for example those commands which only set software options). Another distinctive feature of functions is that they are usually (if they include CPU-intensive operations) processed in separate \textit{threads}, so that the interface remains responsive. In Example 3, the use of some functions (returning different data classes) and commands is shown.

\vspace{2mm}
\small
\begin{mycode}{3}
log( 2 )                      \\ function -> number
append([1, -2], 5)            \\ function -> vector
invert({[1, 3, -1, 4], 2, 2}) \\ function -> matrix
output_precision 8            \\ command -> no output
help invert                   \\ command -> output
\end{mycode} 
\normalsize
\vspace{2mm}

\section{Data types}
\label{datatypes}

BOAT handles four \textit{data classes}: \textit{numbers}, \textit{complex numbers}, \textit{vectors}, and \textit{matrices}. The memory allocation for each data class is automatically determined on the basis of the provided data and the chosen precision. This means that, for example, BOAT will automatically decide whether to define a vector of \textit{integers} or \textit{floats}. Examples of different data class definitions (and assignments to variables) is shown in Example 4.

\vspace{2mm}
\small
\begin{mycode}{4}
$int_var = 54987   // integer number
$real_var = exp(1) // real number 
$com_var = {1, -3} // complex number
$vec_var = [1, 3.1415, 0] // vector
$mat_var = {[4, 12, -1, 0], 2, 2} //matrix
\end{mycode} 
\normalsize
\vspace{2mm}%$

In the following an overview of the supported data classes is provided. Detailed information and the full list of supported operations and functions is available at online user guide \cite{BOATWEB}.

\subsection{Numbers}

As it will be discussed in more detail in Section \ref{arbprec}, BOAT supports numbers with arbitrary precision. By default numbers are internally handled with 256 bit precision and displayed with 8 digits of precision (compatible with 32 bit precision). The precision of 256 bit is very optimized in BOAT and it is the recommended value to use. Operations involving integer numbers are automatically identified and a proper cast to integer in internally performed.

\subsection{Complex Numbers}

A complex number is defined by enclosing its real and imaginary parts, separated by a comma, in curly brackets, as shown in the following example.

\vspace{2mm}
\small
\begin{center}
\begin{minipage}{0.52\textwidth}
\begin{mycode}{5}
{2, -3} 
$x = {2, -1}
$y = {-1, 3}
im($x*$y)
\end{mycode} 
\end{minipage}%
\hspace{4mm}%
\begin{minipage}{0.38\textwidth}
\begin{myout}{5}
2 - i 3


7
\end{myout} 
\end{minipage}
\end{center}
\normalsize
\vspace{2mm}%$

The fourth expression in Example 4 uses the \textit{function} \verb|im| to get the imaginary part of a complex number. 

\subsection{Vectors}

A vector is a sequence of data elements of the same data type. It is defined by enclosing its components, separated by a comma, within square brackets. The data type of the vector components is automatically determined by BOAT. The following example shows the syntax to define a  vector, the assignment to a variable, and few simple operations with vectors and numbers.

\vspace{2mm}
\small
\begin{center}
\begin{minipage}{0.51\textwidth}
\begin{mycode}{6}
[1, 2, 3] 
$x = [ -1, log( 2 )] 
$x
$x + [1, 2] - 10
dotprod([1, -2],[-3, 4]) 
\end{mycode} 
\end{minipage}%
\hspace{4mm}%
\begin{minipage}{0.44\textwidth}
\begin{myout}{5}
[1, 2, 3]

[-1, 0.69314718]
[-10, -7.3068528]
-11
\end{myout} 
\end{minipage}
\end{center}
\normalsize
\vspace{2mm}%$

Example 4 also shows that \textit{functions} returning numbers (\verb|log| in the example) can also be used in the vector definition.

\subsection{Matrices}

Matrices are defined by curly brackets, including a vector of data and two integers, \verb|r| and \verb|c|, specifying the number of rows and columns of the matrix respectively. As for vectors, the data type of the matrix elements (and the consequent memory allocation for the matrix data class) is automatically determined by BOAT. In the following example a $2\times3$ and a $3\times3$ matrix is defined.

\vspace{2mm}
\small
\begin{center}
\begin{mycode}{7}
{ [1, 3.4, 21.6, 19, -0.1, 10], 2, 3 }
{ [1, 3.4, 21.6, 19, -0.1, 10], 3, 3 } 
\end{mycode} 
\begin{myout}{7}
  [  1   3.4  21.6 
    19  -0.1   10  ]
 
  [  1   3.4  21.6 
    19  -0.1   10 
     0   0      0  ]
\end{myout} 
\end{center}
\normalsize
\vspace{2mm}%$

As shown in Example 6, if the declared size of the matrix is bigger than the size of the vector data, the remaining elements are set to 0. BOAT has full support for matrix arithmetic and provides a set of functions for the most common operations (determinant, inverse, trace, etc...). The full list is available in the user guide \cite{BOATWEB}.

\subsection{Dataset}

\begin{figure}[H]
  \centering
  \caption{The graphical interface to import and edit datasets in BOAT.}
  \includegraphics[width=\textwidth]%
    {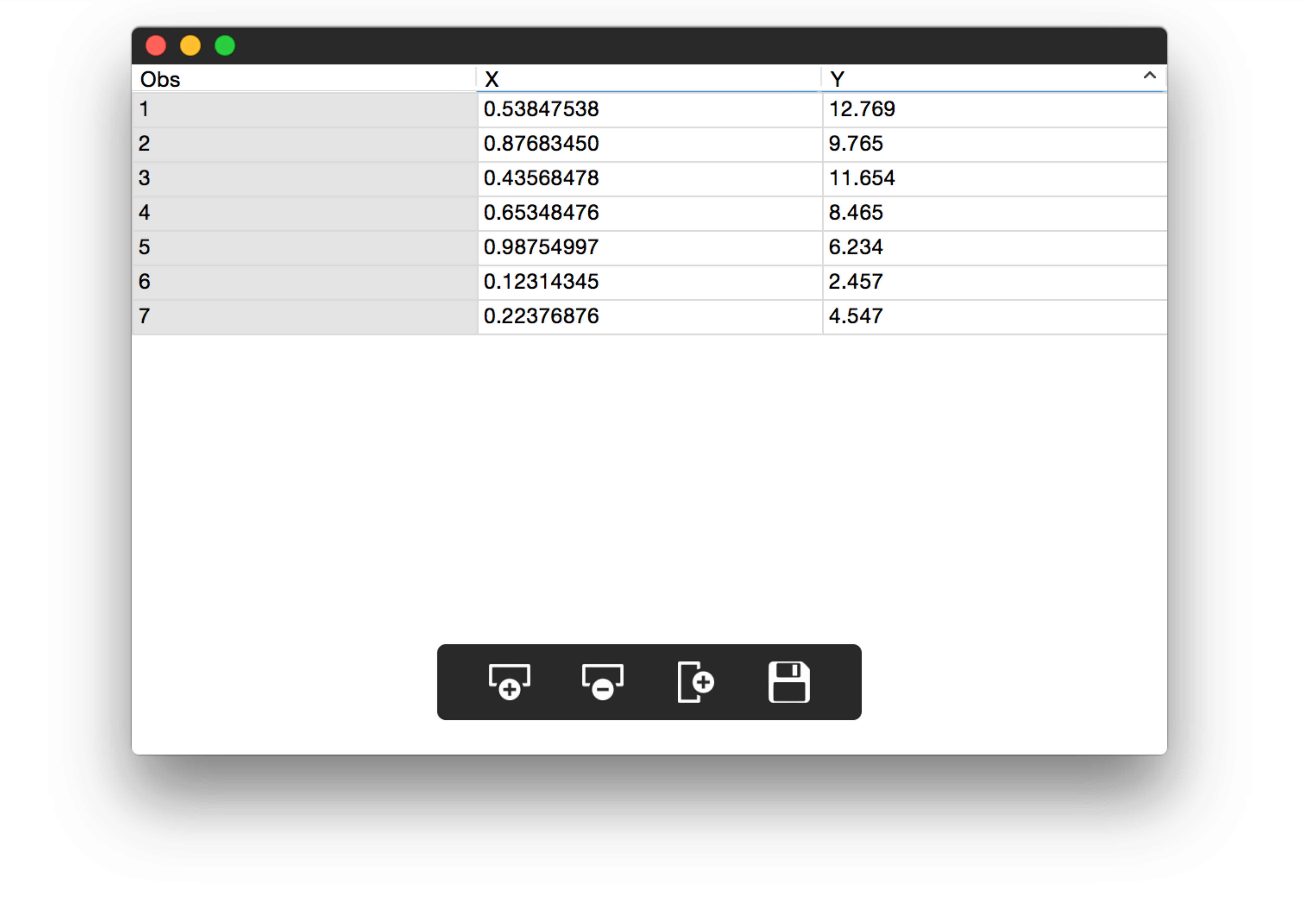}
\label{f:GUIFig2}
\end{figure}

In addition to the four data classes described before, the \textit{dataset} is a convenient object to handle input data. As already introduced in Section \ref{infopanel}, a dataset is a collection of data which are organized as a table of observations (rows) and variables (columns). BOAT provides a handy graphical interface to import and edit datasets, which is shown in Figure \ref{f:GUIFig2}.

Datasets can be imported from few formats: CSV files, tabulated text files and xls files. Variables from a dataset can be accessed by specifying the corresponding column index (0-based) in the dataset table. For instance, given the dataset shown in Figure \ref{f:GUIFig2} called \verb|mydataset|, the statement in Example 7 accesses the variable \verb|X|.

\vspace{2mm}
\small
\begin{center}
\begin{minipage}{0.33\textwidth}
\begin{mycode}{8}
$mydataset[0]
\end{mycode} 
\end{minipage}%
\hspace{4mm}%
\begin{minipage}{0.635\textwidth}
\begin{myout}{8}
[0.6439767, 0.0746277, 0.79...
\end{myout} 
\end{minipage}
\end{center}
\normalsize
\vspace{2mm}%$

\section{Arbitrary Precision}
\label{arbprec}

As already introduced in Section \ref{introduction}, BOAT supports \textit{arbitrary precision arithmetic}, which means that calculations can be performed with a precision which is only limited by the amount of the available memory of the host system. Unlike other available solutions, BOAT not only supports arbitrary precision for integer and float numbers, but for all data classes defined in Section \ref{datatypes}. This means that also vector and matrix calculations can benefit from high precision.

Handling numbers with many digits of precision comes at the price of CPU time, which for most of the calculations is not justified. For this reason, BOAT allows to set the precision before each calculation. The \textit{command} \verb|precision| sets the internal precision to handle float numbers. The precision is specified by passing the number of 32 bit words used to store the number. If no value is specified, a summary of the current precision scheme is returned to the output console.

\vspace{2mm}
\small
\begin{center}
\begin{mycode}{9}
precision 2
precision
\end{mycode} 
\begin{myout}{9}

  Internal precision is set to 2 (memory blocks)
  Actual precision: 64 bits
  Number of printed digits: 16 
\end{myout} 
\end{center}
\normalsize
\vspace{2mm}%$

In the following example, the same calculation is performed with two different levels of precision. In the first case the number $\pi$ (\verb|pi()|) is computed with 64 bit precision (\verb|precision 2|), while, in the second case, a 192 bit precision (\verb|precision 6|) is used. The function \verb|pi()| uses the Bailey-Borwein-Plouffe  formula for calculating $\pi$ \cite{piArt}.

\vspace{2mm}
\small
\begin{center}
\begin{mycode}{10}
precision 2
pi()
precision 6
pi()
precision
\end{mycode} 
\begin{myout}{10}
3.141592653589793
3.14159265358979323846264338327950288419716939938

  Internal precision is set to 6 (memory blocks)
  Actual precision: 192 bits
  Number of printed digits: 48
\end{myout} 
\end{center}
\normalsize
\vspace{2mm}%$

In some cases, it could be preferable to work with many digits of precision but to reduce the number of digits displayed on screen or stored into a report. The command (\verb|output_precision|) specifies the number of digits to be printed. In the following example the calculation of \verb|log(2)| is performed with 192 bit precision but the result is printed with with just two significant figures.

\vspace{2mm}
\small
\begin{center}
\begin{mycode}{11}
precision 6
output_precision 2
precision
pi() 
\end{mycode} 
\begin{myout}{11}

     Internal precision is set to 6 (memory blocks)
     Actual precision: 192 bits
     Number of printed digits: 2
     
  0.69
\end{myout}       
\end{center}
\normalsize
\vspace{2mm}%$

\section{Charts}

BOAT comes with a chart suite for data visualization. The list of supported charts is available at the online user guide \cite{BOATWEB}. The function \verb|plot| provides a general template to plot data, starting from vectors or datasets. The following example shows the creation of a plot starting from two same-size vectors.

\vspace{2mm}
\small
\begin{center}
\begin{mycode}{12}
$x = sequence(-1, 1, 0.1)
$y = cos( $x ) * sin( $x ) 
plot($x, $y, xtitle="x [rad]", ytitle="cos(x)*sin(x)")
\end{mycode} 
\end{center}
\normalsize
\vspace{2mm}%$

\begin{figure}[H]
  \centering
  \caption{Chart windows in BOAT. The chart inspector on the left allows to change the appearance of a chart.}
  \includegraphics[width=0.8\textwidth]%
    {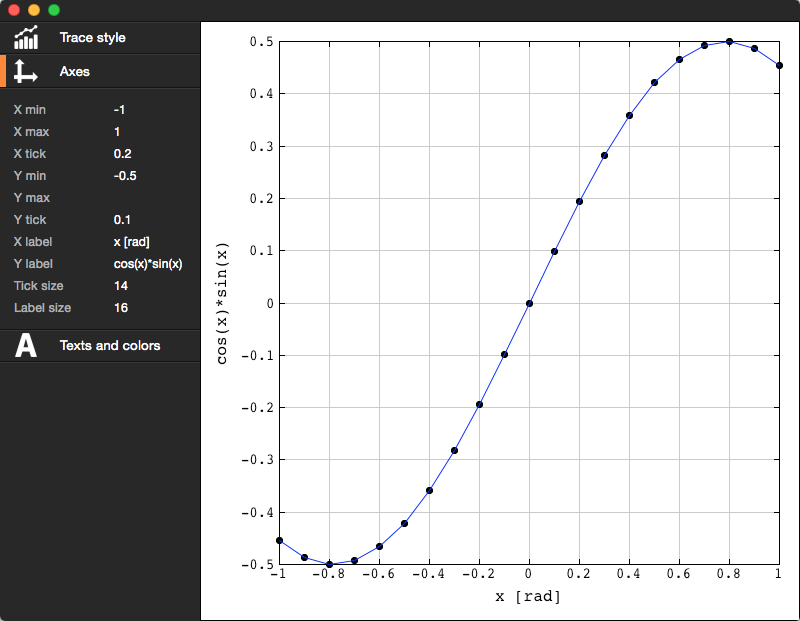}
\label{plotImage}
\end{figure}

The first assignment statement  uses the function \verb|sequence| to create a vector with values from \verb|-1| to \verb|1| with a step of \verb|0.1|. In the second statement functions \verb|cos| and \verb|sin| act on each component of the vector \verb|x| to create a new vector \verb|y|. The output of Example 12 is shown in Figure \ref{plotImage}.

The layout of a chart can be customized either by specifying the options in the corresponding statement (as it was done with \verb|xtitle| and \verb|ytitle| in Example 12) or by using the \textit{chart inspector panel}. The chart inspector panel (the left panel in Figure \ref{plotImage}), can be optionally shown from each chart to edit its appearance. 

Although \verb|plot| is the general template to produce plots, some other functions (like \verb|frequency| for example) automatically produce a chart as output. 

\section{Reports}

Some functions and commands can optionally produce printable reports, which usually contain charts, tables and text. All the reports, produced by a project, are accessible from the corresponding tab in the \textit{info panel}. At present the supported formats are: images, RTF (for compatibility with almost all word processors), and HTML.

\begin{figure}[H]
\label{reportImage}
  \centering
  \caption{HTML report from a one-sample $z$-test (two-sided).}
  \includegraphics[width=0.8\textwidth]%
    {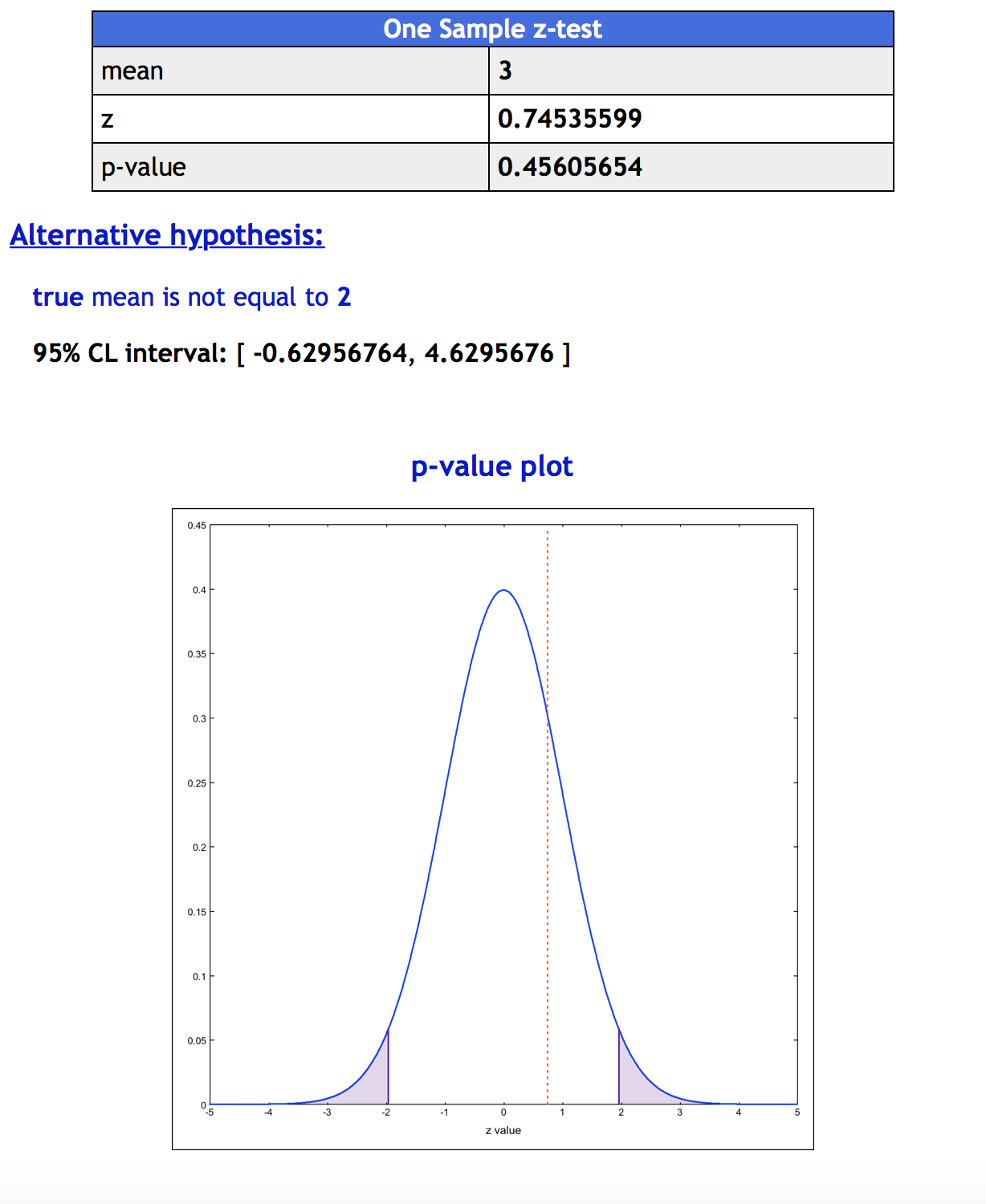}
\end{figure}

The following example performs a one-sample z-test on the vector (\verb|$x|). The option \verb|report=true| enables the creation of the HTML report shown in Figure \ref{reportImage}, which summarizes all the relevant information from the test.

\vspace{2mm}
\small
\begin{center}
\begin{mycode}{12}
$x = [9, 3, -1, -2, 4, 5]
ztest( $x, 2, 3, report=true )
\end{mycode} 
\end{center}
\normalsize
\vspace{2mm}%$

\section{Conclusions}

BOAT is a cross-platform software for data analysis and numerical computing, which can be freely downloaded from the official web page \cite{BOATWEB}. It has been designed and developed with the goal of providing an easy-to-use tool for statistical data analysis, with support for arbitrary precision. The very first release of the software basically laid the foundation, but expansion of features and optimization are expected for the next versions.

\bibliography{mybibfile}

\end{document}